\documentstyle[aps,twocolumn,amssymb,epsf]{revtex}
\begin{document}

\title{
 Thermodynamics of Deconfined Matter at Finite Chemical Potential
 in a Quasiparticle Description
}

\author{{\sc
 A.~Peshier$^{1,2}$, B.~K\"ampfer$^1$, G.~Soff$^2$}}

\address{
 $^1$Forschungszentrum Rossendorf, PF 510119, 01314 Dresden, Germany\\
 $^2$Institut f\"ur Theoretische Physik,
     Technische Universit\"at Dresden, 01062 Dresden, Germany}

\date{June 9, 1999}

\maketitle

\begin{abstract}
 An effective quasiparticle description of the thermodynamics
 of deconfined matter, compatible with both finite-temperature
 lattice data and the perturbative limit, is generalized to
 finite chemical potential.
 Within this approach, the available 4-flavor lattice equation of
 state is extended to finite baryon density, and implications for
 cold, charge-neutral deconfined matter in $\beta$-equilibrium in
 compact stars are considered.
 \\
 {\it PACS number(s):} 12.38.Mh, 31.15.Lc
\end{abstract}
\vspace*{5mm}

 The equation of state (EoS) represents an important interrelation of state
 variables describing matter in local thermodynamical equilibrium.
 All microscopic characteristics are integrated out and only the macroscopic
 response to changes of the state variables as e.\,g., the temperature $T$
 and the chemical potential $\mu$ are retained.
 Via the Gibbs equation, $e = T s + \mu n - p$, the energy density $e$ is
 related to the entropy and particle densities, $s = \partial p / \partial T$
 and $n = \partial p / \partial \mu$, respectively, and the pressure $p$.
 The thermodynamical potential $p(T, \mu)$ thus provides all information
 needed to evaluate, e.\,g., sequences of stellar equilibrium configurations
 by means of the Tolman-Oppenheimer-Volkov equation, or the evolution of the
 universe via Friedman's equations, or the dynamics of heavy-ion collisions
 within the framework of relativistic Euler equations.
 In the examples mentioned not only excited hadron matter is of relevance,
 rather, at sufficiently high density or temperature, a plasma state of
 deconfined quarks and gluons is the central issue.

 Quantum Chromodynamics (QCD) is nowadays the generally accepted fundamental
 theory of interacting quarks and gluons. The challenge, therefore, consists
 in the derivation of the EoS of deconfined matter directly from QCD.
 In first attempts \cite{McLerran} the EoS of cold quark matter was derived
 perturbatively up to order ${\cal O}(g^4)$ in the coupling $g$.
 Finite temperatures have been considered \cite{Zhai}, where the perturbative
 expansion is extended up to the order ${\cal O}(g^5)$.
 However, in the physically relevant region the coupling is large, so
 perturbative methods seem basically to fail and, consequently,
 nonperturbative evaluations are needed.
 Present lattice QCD calculations can accomplish this task, and indeed the
 EoS of the pure gluon plasma \cite{EoS_g} and of systems containing two or
 four light dynamical quark flavors \cite{EoS_2f,EoS_4f} are known at finite
 temperature, and simulations with physical quark masses may be available
 in the near future.

 As a matter of fact, current QCD lattice calculations are restricted to
 the chemical potential $\mu = 0$ while a more detailed understanding of,
 e.\,g., the structure of quark cores in massive neutron stars, the baryon
 contrast prior to cosmic confinement, or the evolution of the baryon charge
 in the midrapidity region of central heavy-ion collisions requires the EoS
 at vanishing net quark density.
 Here we are going to develop an EoS, based on a phenomenological
 quasiparticle model, which extrapolates available lattice results
 at $\mu=0$ to finite values of $\mu$ (even at $T = 0$) and at the
 same time interpolates smoothly to the asymptotic regime of QCD at
 $T \to \infty$ and $\mu \to \infty$, with the strangeness degree of
 freedom properly included.
 With our model we focus on deconfined matter and do not touch upon
 problems of confined matter or the deconfinement transition itself;
 the latter might be covered via usual {\it ad hoc} procedures.

 We consider an $SU(N_c)$ plasma of gluons, with $N_c=3$ for QCD, and
 $N_f$ quark flavors in thermodynamic equilibrium.
 Within the approach outlined below, the interacting plasma is described
 in terms of a quasiparticle system, a picture arising asymptotically from
 the in-medium properties of the constituents of the plasma.
 For thermal momenta $k \ge {\cal O}(T)$ the relevant modes, transversal
 gluons and quark particle excitations, propagate predominantly on-shell
 with dispersion relations $\omega_i^2(k) \approx m_i^2 + k^2$ and
 \begin{equation}
  m_i^2
  =
  m_{0i}^2 + \Pi_i^* \, ,
  \label{m_eff}
 \end{equation}
 while the longitudinal gluonic and helicity flipped quark states are
 essentially unpopulated \cite{LeBellac}. Neglecting sub-leading effects,
 $\Pi_i^*$ are the leading order on-shell selfenergies of parton species $i$.
 Depending on the coupling $G^2$, the temperature and chemical potential
 as well as the rest mass $m_{0i}$ (the latter vanishing for gluons), the
 $\Pi_i^*$ are given by the asymptotic values of the hard thermal/density
 loop selfenergies \cite{LeBellac},
 \begin{eqnarray}
  \Pi_q^*
  &=&
  2\, \omega_{q0}\, (m_0 + \omega_{q0}) \, ,
  \quad
  \omega_{q0}^2
  =
  \frac{N_c^2-1}{16N_c}\, \left[ T^2+\frac{\mu_q^2}{\pi^2} \right] G^2 \, ,
  \nonumber \\
  \Pi_g^*
  &=&
  \frac16
  \left[
     \left( N_c+\frac12\, N_f \right) T^2
   + \frac{N_c}{2\pi^2} \sum_q \mu_q^2
  \right] G^2 \, .
 \label{Pi}
 \end{eqnarray}
 Generalizing the approach of ref.~\cite{Gorenstein} to a finite chemical
 potential $\mu$ controlling a conserved particle number, the pressure of
 the system can be decomposed into the contributions $p_j$ of the
 quasiparticles, and their mean field interaction $B$,
 \begin{equation}
  p(T,\mu; m_{0i}^2)
  =
  \sum_j p_j(T, \mu_j(\mu); m_j^2) - B(\Pi_i^*) \, ,
  \label{p_eff}
 \end{equation}
 where formally $p_j$ is the pressure of an ideal gas of bosons or fermions
 with state dependent effective masses (\ref{m_eff},\ref{Pi}).
 By the stationarity of the thermodynamic potential $\Omega = -pV$ under
 functional variation of the selfenergies \cite{LeeYang}, which in the
 present approach simplifies to $\partial p / \partial \Pi_j^* = 0$, $B$
 is related to the quasiparticle masses,
 \begin{equation}
   \frac{\partial B}{\partial \Pi_j^*}
   =
   \frac{\partial p_j(T,\mu_j; m_j^2)}{\partial m_j^2} \, ,
  \label{stat}
 \end{equation}
 which implies that the entropy and particle densities are given by the
 sum of the quasiparticle contributions
 \begin{equation}
   s_j
   =
   \left.
    \frac{\partial p_j(T, \mu_j; m_j^2)}{\partial T}
   \right|_{m_j^2}
   \! , \;
   n_j
   =
   \left.
    \frac{\partial p_j(T, \mu_j; m_j^2)}{\partial \mu_j}
   \right|_{m_j^2}
   \, .
 \end{equation}
 The quasiparticle approach represents an effective resummation of the
 leading-order thermal contributions \cite{TFT}. Hence, it is expected
 to be an appropriate framework as long as the spectral properties of
 the relevant plasma excitations do not differ qualitatively from their
 asymptotic form.
 In the hot $\phi^4$ theory, for which an equivalent thermodynamic
 quasiparticle description was derived by resumming all tadpole diagrams
 \cite{EL}, this assumption was shown to be fulfilled also at larger
 values of the coupling by resumming the propagator beyond 1-loop order
 \cite{WangHeinz}.
 Corroborating the assumption for QCD thermodynamics at $\mu=0$,
 the quasiparticle approach was found to be in a remarkably good
 agreement with lattice data even close to the confinement region
 \cite{PR,J_Phys,LevaiHeinz}, with the effective coupling $G^2$
 in eq.~(\ref{Pi}) parameterized by
 \begin{equation}
   G^2(T,\mu=0)
   =
   \frac{48\pi^2}
        {\left( 11 N_c - 2\, N_f \right)
         \ln\left(\displaystyle\frac{T+T_s}{T_c/\lambda}\right)^{\!\!2}} \, ,
   \label{G2}
 \end{equation}
 interpolating smoothly to the asymptotic QCD limit.

 For finite values of $\mu$, we observe that due to the stationary property
 (\ref{stat}) the dependence of the function $B$ on the state variables is
 determined by
 \begin{equation}
  \frac{\partial B}{\partial T}
  =
  \sum_j \frac{\partial p_j}{\partial m_j^2}\,
          \frac{\partial \Pi_j^*}{\partial T}
  \equiv
  B_T ,
  \,
  \frac{\partial B}{\partial \mu}
  =
  \sum_j \frac{\partial p_j}{\partial m_j^2}\,
          \frac{\partial \Pi_j^*}{\partial \mu}
  \equiv
  B_\mu \, .
  \label{B_x}
 \end{equation}
 The pressure (\ref{p_eff}) and thus the function $B$ being a potential
 of the state variables $T$ and $\mu$, the functions $B_T$ and $B_\mu$
 have to respect the integrability condition
 \begin{equation}
  \frac{\partial B_T}{\partial \mu} - \frac{\partial B_\mu}{\partial T}
  =
  \sum_j
  \left[
    \frac{\partial n_j}{\partial m_j^2}\,
     \frac{\partial \Pi_j^*}{\partial T}
   -\frac{\partial s_j}{\partial m_j^2}\,
     \frac{\partial \Pi_j^*}{\partial \mu}
  \right]
  =
  0 \, .
  \label{flow}
 \end{equation}
 In the selfenergies, eq.~(\ref{Pi}), the effective coupling itself
 is a function of the state variables, so eq.~(\ref{flow}) represents
 a first order partial differential equation for $G^2(T, \mu)$.
 With $G^2(T, \mu)$ given, e.\,g., at $\mu=0$ by lattice data, this
 flow equation determines the effective coupling and hence by eqs.\
 (\ref{m_eff}-\ref{p_eff}) and (\ref{B_x}) the EoS of the plasma at
 finite temperature and chemical potential.

 As a first example let us now apply the quasiparticle approach to the
 EoS of the QCD plasma with $N_f=4$ light flavors, which is numerically
 known for $\mu=0$ in a restricted interval of $T$.
 From a fit (cf.\ fig.~1 in ref.~\cite{J_Phys}) of the lattice data
 \cite{EoS_4f} we obtain the parameters $\lambda=6.7$ and $T_s/T_c=-0.81$
 of the effective coupling (\ref{G2}), and the gluon and quark quasiparticle
 degrees of freedom $d_g=20.6$ and $d_q=\frac{4N_c N_f}{2(N_c^2-1)}\, d_g$.
 The large value of the parameter $T_s/T_c$ points to the distinct
 nonperturbative behavior at $T \rightarrow T_c$.
 We note that the ratio of the fitted value of $d_g$ to the expected value
 of 16 is of the same order of magnitude as finite lattice size effects in
 the interaction free limit.

 With this at hand we extrapolate the lattice data to finite values of
 $\mu_q = \mu$, corresponding to a finite net quark density.
 Eq.~(\ref{flow}) is a quasilinear partial differential equation of the form
 \begin{equation}
  a_T\, \frac{\partial G^2}{\partial T}
  +
  a_\mu\, \frac{\partial G^2}{\partial \mu}
  =
  b \, ,
  \label{Nf4flow}
 \end{equation}
 with the coefficients $a_{T,\mu}$ and $b$ depending on $T$, $\mu$ and
 $G^2$, which can be solved by the method of characteristics.
 It is instructive to consider first the asymptotic limit, $m_i^2/T^2 \sim
 G^2 \rightarrow g^2 \rightarrow 0$, where (\ref{Nf4flow}) reduces to
 \begin{equation}
  \pi^2 \left( c T^2 + \frac{\mu^2}{\pi^2} \right)
  \frac1\mu\, \frac{\partial g^2}{\partial \mu}
  -
  \left( T^2 + \frac{\mu^2}{\pi^2} \right)
  \frac1T\, \frac{\partial g^2}{\partial T}
  =
  0 \, ,
 \end{equation}
 with $c = (4 N_c + 5 N_f)/(9 N_f)$.
 This equation yields $g^2 = const$ along the characteristics given by
 $c T^4 + 2 T^2 (\mu/\pi)^2 + (\mu/\pi)^4 = const.$
 As a result of this elliptic flow, the renormalization scale $T_c/\lambda$
 of the effective coupling $g^2(T,0)$ determines the scale $T_c \pi c^{1/4}
 /\lambda$ of $g^2(0,\mu)$, and the pressure (\ref{m_eff}-\ref{p_eff})
 coincides with the perturbative QCD expression up to the order
 ${\cal O}(g^2)$.

\begin{figure}[hbt]
  \epsfysize 8cm
  \centerline{\epsffile{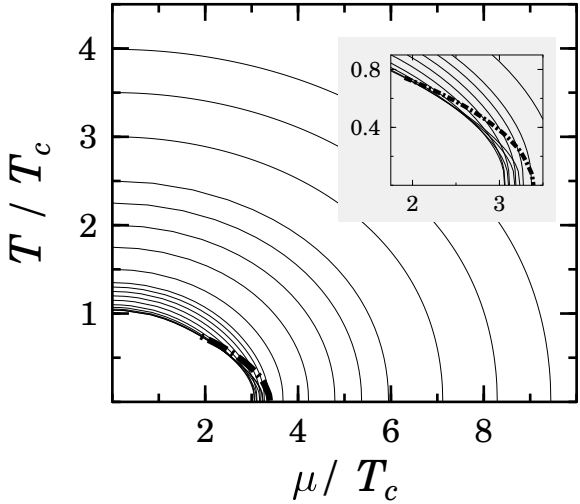}}
  \caption{The characteristics of the coupling flow equation
           (\protect\ref{Nf4flow}) for the
           QCD plasma with $N_f=4$ light flavors for which $G^2(T, \mu=0)$
           is obtained from lattice data \protect\cite{EoS_4f}.
           At leading order the characteristics are curves of constant
           coupling strength.
           The pressure is negative in the region below the dash-dotted line,
           thus excluding the region of intersecting characteristics.
           \label{F:Nf4flow}}
\end{figure}

\begin{figure}[hbt]
  \epsfxsize 7cm
  \centerline{\epsffile{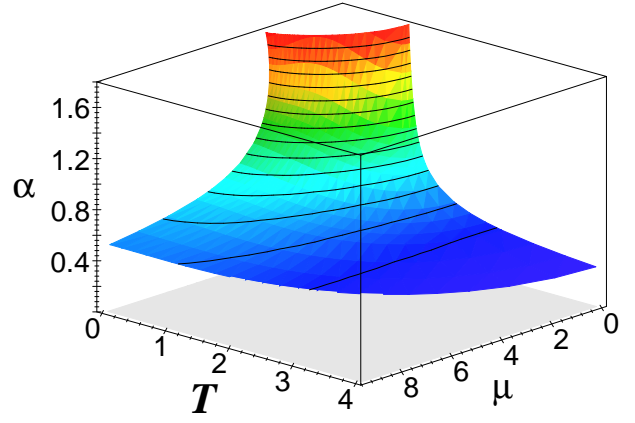}}
  \caption{The effective coupling strength $\alpha = G^2 /(4 \pi)$
           as a function of $\mu$ and $T$
           for the $N_f=4$ plasma in the chiral limit.
           \label{F:Nf4alpha}}
\end{figure}

 The flow (\ref{Nf4flow}) of the effective coupling, with $G^2(T,0)$ obtained
 from the lattice data referred to above, is shown in fig.~\ref{F:Nf4flow}.
 The characteristics related to larger temperatures resemble their
 asymptotic form, and the coupling displayed in fig.~\ref{F:Nf4alpha}
 decreases asymptotically as expected.
 For smaller values of $T$ and $\mu$, the pronounced increase of $G^2$
 is considered to be a sign of the vicinity of the phase transition.
 Indicated by intersecting characteristics, the solution of the flow
 equation is non-unique in a certain low-temperature region.
 In this region, however, the pressure turns out to be negative, so the
 ambiguity is of no physical relevance.
 To obtain the pressure, which is available in tabular form \cite{WEB},
 the function $B(T,\mu)$ can be computed along the characteristics using
 eq.~(\ref{stat}).
 Hereby, the required function $B(T,\mu=0)$ is determined by the first
 equation of (\ref{B_x}) up to an integration constant $B_0 = B(T_c,0)$
 which is fixed by equating the quasiparticle pressure and the lattice
 data at $T_c$.
 For chemical potentials $\mu \sim 2.5\,T_c$ and small temperatures, in
 a region of the phase space corresponding to characteristics emanating
 from the vicinity of the point $(T_c,\mu=0)$, the pressure becomes negative.
 This region of instability provides a lower boundary for the value of the
 confining chemical potential $\mu_c \ge 3.4\, T_c$ at $T=0$.
 As evident from fig.~\ref{F:Nf4flow}, this boundary also excludes the
 region of non-unique flow.
 Hence, the quasiparticle model is intrinsically consistent and therefore
 considered to provide a realistic extension of the EoS obtained by lattice
 calculations to $\mu \neq 0$ even near the confinement transition.

 To conclude this example we remark that by the Feynman-Hellmann relation
 the quasiparticle pressure (\ref{p_eff}) leads to a chiral condensate
 $\langle \bar\psi \psi \rangle \sim m_{0q}$, so for massless quark flavors
 the restoration of chiral symmetry in the deconfined phase is inherent in
 the approach.

 Thermodynamic lattice simulations of QCD with physical quark current masses,
 $m_{0u,d} \sim 0$ for the light flavors and $m_{0s} \sim 150$ MeV for
 strange quarks are still lacking, so the parameters of the quasiparticle
 description cannot be fixed at finite temperature to extend the physically
 relevant EoS to $\mu \neq 0$.
 However, a trial EoS $p^{\rm trial}(T,\mu=0)$ can be constructed by
 reasonable choices of the model parameters until they may be specified
 by precise lattice data.
 Due to the stationarity property (\ref{stat}) of the thermodynamic
 potential, $p^{\rm trial}(T,\mu=0)$ is expected to possess only a weak
 dependence on the parameters.
 The model parameters are restricted to match the hadronic pressure
 $p^{\rm had}$ at the confinement temperature which we assume to be
 $T_c = 150$\,MeV. The uncertainty of $p^{\rm had}(T_c,0)=3.1\cdot10^8\,$
 MeV$^4$ as obtained by a hadron resonance gas model can be absorbed into
 the integration constant $B_0$, which we vary independently beside the
 scaling parameter $\lambda$ in eq.~(\ref{G2}).
 Being related to the QCD scale parameter $\Lambda_{\rm QCD}$, $\lambda$
 is expected to be a slowly increasing function of the number of flavors;
 considering $3 \le \lambda \le 9$ we cover the values $\lambda^{N_f=0} =
 4.2$ as obtained from the pure gluon lattice data \cite{EoS_g} and
 $\lambda^{N_f=4} = 6.7$ in the example above.

\begin{figure}[hbt]
  \epsfysize 8cm
  \centerline{\epsffile{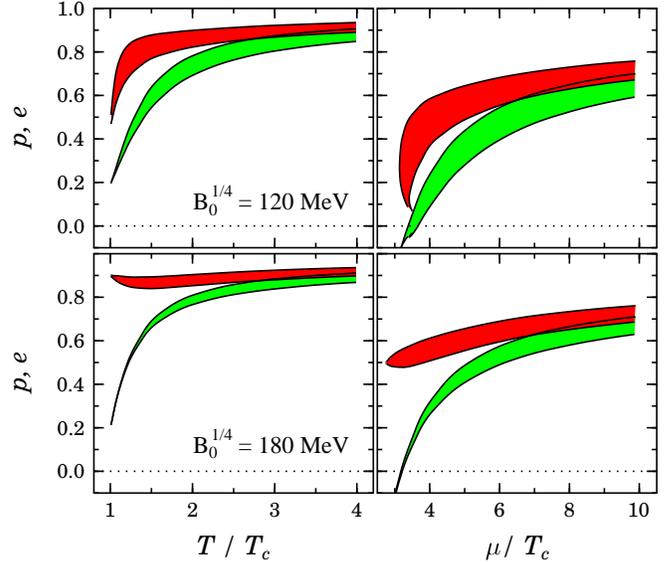}}
  \caption{Total pressure and energy density (lower and upper set of curves,
           respectively) of the charge-neutral, quark-gluon plasma in
           $\beta$-equilibrium, scaled by the values of the free limit.
           The panels on the left (right) show the EoS at $\mu=0\,(T=0)$,
           for values of the model parameter $3 \le \lambda \le 9$
           (lower and upper line, respectively, of hatched area),
           and $B_0^{1/4} = 120\,$MeV (top) and $B_0^{1/4} = 180\,$MeV
           (bottom).
           Non-unique values of the energy only occur in the unphysical
           region, where $p < 0$.
           \label{F:(2+1)eos}}
\end{figure}

 With the aim to study implications for quark matter stars, we consider
 in the following a plasma of gluons, quarks and electrons in
 $\beta$-equilibrium maintained by the reactions $d,s \leftrightarrow
 u + e + \bar\nu_e$, which imply the relations $\mu_d = \mu_s = \mu_u+\mu_e
 \equiv \mu$ among the chemical potentials.
 The electron chemical potential $\mu_e$ as a function of $T$ and $\mu$
 is determined by the requirement of electrical charge neutrality.
 With the electron pressure $p^e$ approximated by its free limit, the
 pressure $p=p^{\rm trial}+p^e$ and the resulting energy density are
 shown for several values of the parameters $\lambda$ and $B_0$ at
 $\mu=0$ in the left panels of fig.~\ref{F:(2+1)eos}.
 Already at temperatures slightly above $T_c$, the scaled energy density
 reaches a saturation-like behavior at some 90\% of the asymptotic limit.
 This feature, known qualitatively from the lattice simulations
 \cite{EoS_g,EoS_2f,EoS_4f}, is to a large extent insensitive to the
 specific choice of the free parameters, while $B_0$ has a distinct
 impact on the latent heat.
 At vanishing temperature, the resulting EoS is displayed in the right
 panels of fig.~\ref{F:(2+1)eos}.
 In this case, the asymptotic values are approached more slowly due to
 the less rapid decrease of the effective coupling with increasing values
 of $\mu$, similar to the $N_f=4$ plasma.

 With the resulting EoS $e(p)$, sequences of hydrostatic equilibrium
 configurations of cold stars can be calculated by the
 Tolman-Oppenheimer-Volkov (TOV) equation (cf.\ eq.~(2.212) of
 ref.~\cite{Glendenning}).
 For energy densities up to several times the nuclear density and at
 temperatures less then some 10 MeV, the EoS of $\beta$-stable quark matter,
 as estimated by our quasiparticle approach can be parameterized by
 $e = 4 \tilde{B} + \tilde\alpha p$.
 While the naive bag model EoS has $\tilde\alpha = 3$, our quasiparticle
 EoS, with the considered choices of the model parameters, yields
 $3.1 \le \tilde \alpha \le 4.5$, indicating the nontrivial nature of the
 interaction.
 For parameters $\lambda \ge 5$, values of $\tilde{B}^{1/4} \ge 200\,$MeV
 are found, while only the extreme choice of $\lambda=3$ allows $\tilde{B}
 ^{1/4}$ as small as 180\,MeV.
 However, the value of $\tilde{B}$, i.\,e., the energy density at small
 pressure, has a strong impact on the star's mass and radius, obtained
 by integrating the TOV equation.
 Compact quark stars of mass $M \ge M_\odot$, e.\,g., could only exist
 for values of $\tilde{B}^{1/4} \le 180...200\,$MeV, depending marginally
 on $\tilde \alpha$.
 We hence conclude that compact stars with typical masses $\sim M_\odot$
 and radii $\sim 10\,$km are unlikely to be composed purely of deconfined
 matter in $\beta$-equilibrium, irrespective of the uncertainty of the
 model parameters of our approach.

 In summary we have generalized a thermodynamic quasiparticle description
 of deconfined matter to finite chemical potential $\mu$ not accessible
 by present lattice calculations.
 Of central importance to the model is the effective coupling $G^2(T,\mu)$
 which can be obtained at $\mu=0$ from available lattice data, proving at
 the same time the applicability of the effective description even close
 to the confinement transition.
 At finite chemical potential, $G^2$ is determined by a flow equation
 resulting from the general requirement of integrability.
 By the flow of the effective coupling, the basic features of the EoS at
 $\mu=0$, namely the nonperturbative behavior near confinement and the
 asymptotics, are mapped into the $T$-$\mu$ plane as exemplified by the
 $N_f=4$ flavor system studied on the lattice.
 An important consequence of the quasiparticle approach is the relation
 of the critical values of temperature and chemical potential.
 For deconfinement matter with physical quark masses, this fact leads to
 the implication that compact stars composed purely of $\beta$-stable
 deconfined matter may be less massive and, hence, more distinct in
 the bulk properties to neutron stars than estimated by other approaches.
 \\[1mm]
 {\bf Acknowledgments:}
 We are grateful to E.~Grosse and F.~Thielemann for initiating the present
 work. The stimulating interest of O.\,P.~Pavlenko is acknowledged.
 The work is supported by BMBF 06DR829/1.

\begin{thebibliography}{99} \newpage
\bibitem{McLerran}
   B.\,A.~Freedman, L.\,D.~McLerran,
   Phys.\ Rev.\ D16, 1130, 1147, 1169 (1977)
\bibitem{Zhai}
   C.~Zhai, B.~Kastening,
   Phys.\ Rev.\ D52, 7232 (1995)
\bibitem{EoS_g}
   G.~Boyd, J.~Engels, F.~Karsch, E.~Laermann, C.~Legeland,
   M.~L\"utgemeier, B.~Petersson,
   Nucl.\ Phys.\ B469, 419 (1996)
\bibitem{EoS_2f}
   C.~Bernard et al.,
   Phys.\ Rev.\ D55, 6861 (1997)
\bibitem{EoS_4f}
   J.~Engels, R.~Joswig, F.~Karsch, E.~Laermann, M.\ L\"utgemeier,
   B.~Petersson,
   Phys.\ Lett.\ B396, 210 (1997)
\bibitem{LeBellac}
   M.~Le Bellac, {\em Thermal Field Theory},
   Cambridge University Press, Cambridge (1996)
\bibitem{Gorenstein}
   M.\,I.~Gorenstein, S.\,N.~Yang,
   Phys.\ Rev.\ D52, 5206 (1995)
\bibitem{LeeYang}
   T.\,D.~Lee, C.\,N.~Yang,
   Phys.\ Rev.\ 117, 22 (1960)
\bibitem{TFT}
   A.~Peshier,
   TFT'98 Proceedings, hep-ph/9809379
\bibitem{EL}
   A.~Peshier, B.~K\"ampfer, G.~Soff, O.\,P.~Pavlenko,
   Europhys.\ Lett.\ 43, 381 (1998)
\bibitem{WangHeinz}
   E.~Wang, U.~Heinz,
   Phys.\ Rev.\ D53, 899 (1996)
\bibitem{PR}
   A.~Peshier, B.~K\"ampfer, O.\,P.~Pavlenko, G.~Soff,
   Phys.\ Rev.\ D54, 2399 (1996)
\bibitem{J_Phys}
   B.~K\"ampfer, O.\,P.~Pavlenko, A.~Peshier, M.~Hentschel, G.~Soff,
   J.\ Phys.\ G23, 2001c (1997)
\bibitem{LevaiHeinz}
   P.~Levai, U.~Heinz,
   Phys.\ Rev.\ C57, 1879 (1998)
\bibitem{WEB}
   data files available from \mbox{http://www.fz-rossendorf.de/}
   FWK/MITARB/Peshier
\bibitem{Glendenning}
   N.\,K.~Glendenning,
   {\em Compact stars}, Springer Verlag, New York (1997)
\end{thebibliography}
\end{document}